\documentclass[
 reprint,
 amsmath,amssymb,
 nofootinbib,
 aps,
 prl,
 superscriptaddress,
]{revtex4-1}
\usepackage{aas_macros}
\usepackage{graphicx}
\usepackage{dcolumn}
\usepackage{bm}
\usepackage{amsmath}
\usepackage{float}
\usepackage{lipsum}
\usepackage{color}
\usepackage{times}
\usepackage{textcomp}
\usepackage{latexsym}
\usepackage[hidelinks]{hyperref}
\usepackage{acro}
\usepackage{mathtools}
\usepackage{url}
\usepackage{aas_macros}
\usepackage[utf8]{inputenc}
\usepackage{physics}

\begin{document}

\title{Extreme Dark Matter Tests with Extreme Mass Ratio Inspirals}

\author{Otto A. Hannuksela}
\email{hannuksela@phy.cuhk.edu.hk}
\thanks{{\footnotesize \href{https://orcid.org/0000-0002-3887-7137}{orcid.org/0000-0002-3887-7137}}}
\affiliation{
Department of Physics, The Chinese University of Hong Kong, Sha Tin, N.T., Hong Kong}

\author{Kenny C. Y. Ng}
\email{chun-yu.ng@weizmann.ac.il}
\thanks{{\footnotesize \href{http://orcid.org/0000-0001-8016-2170}{orcid.org/0000-0001-8016-2170}}}
\affiliation{
Department of Particle Physics and Astrophysics, Weizmann Institute of Science, Rehovot, Israel}

\author{Tjonnie G. F. Li}
\email{tgfli@cuhk.edu.hk}
\thanks{{\footnotesize \href{http://orcid.org/0000-0003-4297-7365}{orcid.org/0000-0003-4297-7365}}}
\affiliation{
Department of Physics, The Chinese University of Hong Kong, Sha Tin, N.T., Hong Kong}

\date{June 27, 2019}

\begin{abstract}
\noindent
Future space-based laser interferometry experiments such as LISA are expected to detect $\cal O$(100--1000) stellar-mass compact objects (e.g., black holes, neutron stars) falling into massive black holes in the centers of galaxies, the so-called extreme-mass-ratio inspirals~(EMRIs). If dark matter forms a ``spike'' due to the growth of the massive black hole, it will induce a gravitational drag on the inspiraling object, changing its orbit and gravitational-wave signal. 
We show that detection of even a single dark matter spike from the EMRIs will severely constrain several popular dark matter candidates, such as ultralight bosons, keV fermions, MeV--TeV self-annihilating dark matter, and sub-solar mass primordial black holes, as these candidates would flatten the spikes through various mechanisms. Future space gravitational wave experiments could thus have a significant impact on the particle identification of dark matter. 
\end{abstract}

\maketitle

\section{Introduction}
\noindent

Astrophysical and cosmological observations from vastly different scales have established the existence of non-baryonic substance---dark matter~(DM)--- that makes up around 85\% of all known matter~\cite{bertone2005particle, bertone2018new}. Significant effort has gone into identifying DM, whose discovery will be a crucial breakthrough in fundamental physics and understandings of the Universe. 

Many probes of DM utilize regions of high DM densities, such as the Galactic halo and dwarf spheroidals~(dSphs). One attractive target for such searches is the DM spikes, dense concentrations of DM surrounding massive black holes~(BH) in centers of galaxies~\cite{gondolo_dark_1999,sadeghian_dark_2013, Ferrer:2017xwm}, formed as the BHs grow adiabatically.  If DM self-annihilates, the spikes would significantly boost the annihilation rate. This has led to searches of bright isolated gamma-ray sources in the sky as well as constraints in DM annihilation cross sections~\cite{merritt_dark_2002, Aharonian:2008wt,Bertone:2009kj, Bringmann:2009ip, Lacki:2010zf,banados_emergent_2011, Fields:2014pia, Shelton:2015aqa, lacroix_ruling_2015, wanders_no_2015, sandick_black_2016,shapiro_weak_2016}. 
However, the abundance of the BHs and the DM spike properties~(notably the density slope) are uncertain~\cite{ullio_dark-matter_2001,merritt_dark_2002,merritt2004evolution,bertone_time-dependent_2005}. 
The sub-parsec regions of the BHs can only be probed in a few selected systems~\cite{lacroix_unique_2016, Bar:2019pnz, Davoudiasl:2019nlo}. 
Furthermore, there is no guarantee that DM can self-annihilate. 
This dual uncertainty makes it difficult to constrain either the DM spikes or DM particle properties robustly.  

Thankfully, this picture could change dramatically soon. 
With the advent of gravitational wave (GW) astronomy, future space laser interferometry experiments, such as LISA~\cite{2017arXiv170200786A}, Taiji~\cite{Guo:2018npi}, and TianQin~\cite{Luo:2015ght} can detect hundreds to thousands of stellar-mass compact objects~(BHs, neutrinos stars, and white dwarfs) falling into intermediate or supermassive BHs~\cite{Babak:2017tow}. 
These events are called extreme mass ratio inspirals (EMRIs). 
GWs from EMRIs can probe the properties of the surrounding DM spikes~\cite{eda2013new, eda2015gravitational, barausse2015environmental, yue2018gravitational,hannuksela2019probing}, or similar concentrations~\cite{hannuksela2019probing}.
By measuring the DM spike profile using purely gravitational interactions, it is possible to reliably detect the spike and then simultaneously constrain the properties of DM. 

Previously, it was demonstrated that EMRI measurements could be used to infer ultralight boson properties~\cite{hannuksela2019probing}. 
The potential to test DM properties were also briefly noted in Refs.~\cite{eda2013new, eda2015gravitational, bertone2018new}. 
However, a clear picture of the implications of EMRI detection to various DM models is absent. 

In this Letter, we show that even with a single GW detection of DM spike, one can place strong constraints on the properties of several popular DM models. 
These include ultralight bosons, keV fermions, self-annihilating DM, and primordial BHs.
We provide the principal arguments and order-of-magnitude, but robust, determination of the constraints.

\section{DM spike formation around BHs\label{sec:dmspike} }

We consider the massive BHs that sit at the centers of DM halos. The density profile near the BH is approximated by 
\begin{equation} \label{eq:halorho}
 \rho(r) \simeq \rho_0 (r/r_0)^{-\gamma} \, .
\end{equation}
When the BH grows adiabatically, the surrounding DM density also changes according to the changes in the gravitational pull and forms a spike~\cite{gondolo_dark_1999,sadeghian_dark_2013,nishikawa2017primordial}: 
\begin{equation} \label{eq:dmdistribution}
 \rho_{\rm sp}(r) = \rho_R (1 - 8M_{\rm BH}/r)^{3} (r/R_{\rm sp})^{-\alpha}\, ,
\end{equation}
where $\alpha = (9-2\gamma)/(4-\gamma)$, which yields $2.25<\alpha<2.5$ for $0<\gamma<2$ (here and below we will use units of $G = c = \hbar = 1$). 
The factors $\rho_R$ and $R_{\rm sp}$ depend on $\gamma$ and the BH mass $M_{\rm BH}$ through the $M$--$\sigma$ relation~(\cite{nishikawa2017primordial}, and see Appendix). $\gamma = 1$ corresponds to the NFW profile, and we consider $\gamma = 2$ to be the optimistic case. 

We note that depending on the merger history and the stellar environment of the BHs, the DM spikes could be disrupted and end up with shallower final slopes~\cite{nakano1999cusp,ullio_dark-matter_2001, merritt_dark_2002, bertone2005dark, vasiliev2008dark}. However, the extent of these effects is still being debated~\cite{Fields:2014pia, Shelton:2015aqa}, and it has been suggested that spikes around lighter BHs are less affected by mergers~\cite{zhao2005dark,bertone2005new}. 
We emphasize that the EMRI GW detection itself does not depend on the formation scenarios or properties of DM~(e.g., annihilation), and thus provides a model-independent test for the spikes.

Figure~\ref{fig:densityrecovery} shows the peak spike density $\rho_{\rm peak} \approx \rho_{\rm sp}(20M_{\rm BH})$ for $M_{\rm BH} \in [10^3, 10^6]M_{\odot}$. For $\gamma =1$, $\rho_{\rm peak}$ can reach up to $10^{23}\,{\rm GeV/cm^{2}}$.  For the optimistic $\gamma = 2$ scenario, $\rho_{\rm peak}$ can reach up to $10^{26}\,{\rm GeV/cm^{2}}$. If detected, the DM spikes will be the most compact DM structures ever known.

\section{DM spike detection with EMRI}
\noindent

We now turn to the detection prospects of DM spikes with GWs from EMRI events. 

\noindent\textbf{\underline{Detecting the DM Spikes}} 
We model GWs from extreme mass-ratio binary systems interacting with DM spikes (following Refs.~\cite{eda2013new, eda2015gravitational}), and recover the spike parameters in LISA setting.
As the compact objects fall into the BHs, they lose energy and change orbits due to dynamical friction. The properties of the DM spike are thus encoded in the emitted GWs. 

Our set-up is as follows: 
For each BH mass, $M_{\rm BH}$, we set a constant signal-to-noise ratio of ${\rm SNR}=30$ (the usual detection threshold for EMRIs~\cite{gair2004event}). 
Dark matter effects are modeled at the lowest post-Newtonian (PN) order for $\rho_{\rm dm}(r) = \rho_{\rm peak} (r/20M_{\rm BH})^{-\alpha}$ profile (as in Ref.~\cite[][]{eda2015gravitational}). 
Other binary interactions are at 2.5 PN~\cite{Berti:2004bd}. 
We set the mass ratio of the binary $q=\mu/M_{\rm BH}=10^{-4}$, and we assume LISA sensitivity with angle-averaged antenna pattern functions~\cite{Cornish:2018dyw}. 
We then recover the parameters of the injected waveforms, including $\rho_{\rm peak}$ and the density slope $\alpha$, by the Fisher Information Matrix (FIM) method~\cite{Berti:2004bd}\footnote{Specifically, we recover the parameters $\vec{\theta} \in \{ A, \phi_c, t_c, \log \mathcal{M}_c, \log \eta, \beta, \sigma, \rho_{\rm peak}, \alpha \}$, where $\beta$, $\sigma$ represent spin-orbit and spin-spin contributions to the phasing~\cite{Berti:2004bd} }. 
We assume $5$ years of orbital time and the last orbital cycle at $r=20M_{\rm BH}$.  

\begin{figure}
 \includegraphics[width=\columnwidth]{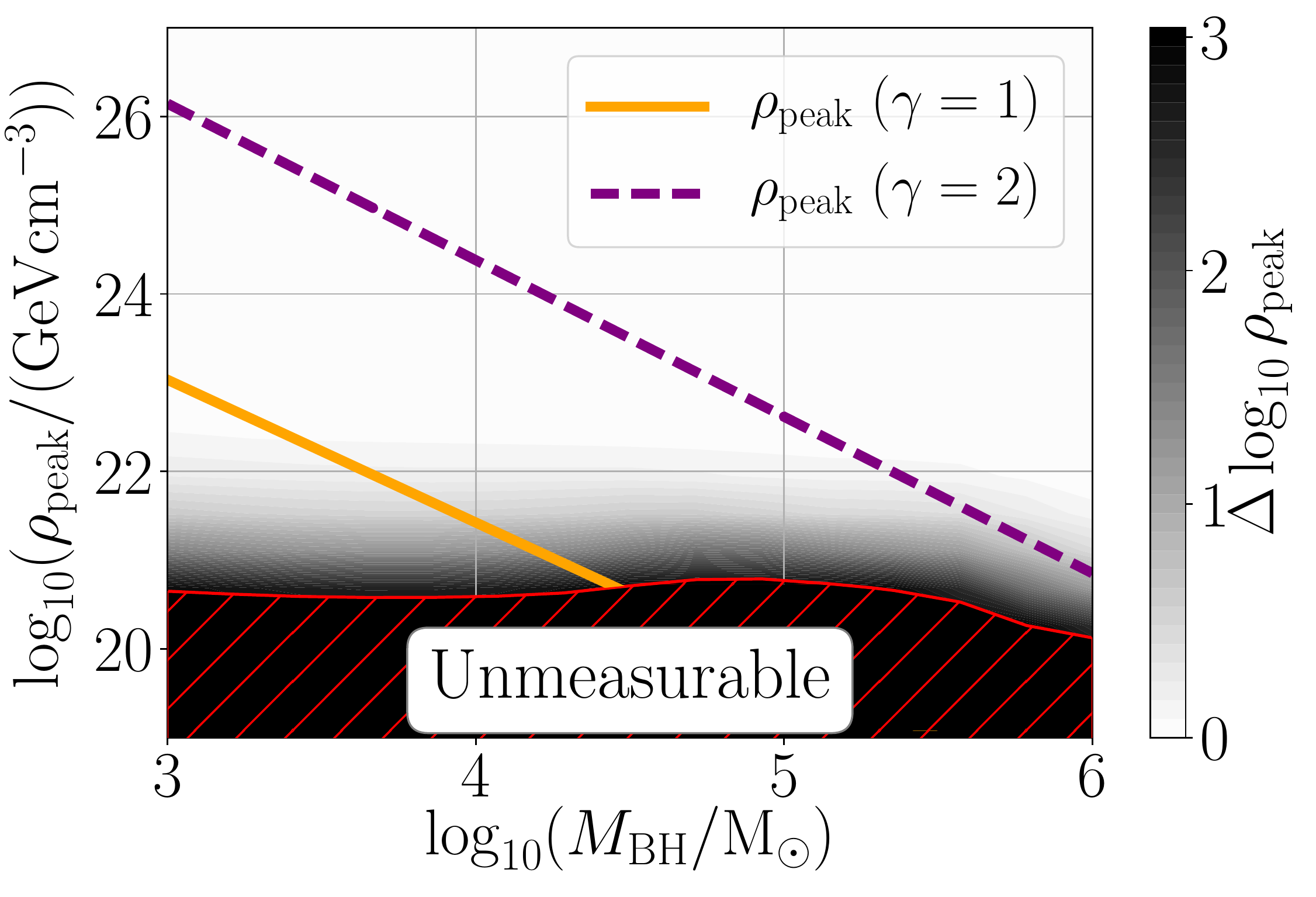}
 \caption{
Uncertainty in the peak DM spike density $\Delta \log_{10} \rho_{\rm peak}$ from EMRI GW measurement around a massive BH~(grey scale) as a function of $\rho_{\rm peak}$ and BH mass $M_{\rm BH}$.
We set the mass ratio $q=10^{-4}$ and signal-to-noise ratio ${\rm SNR}=30$. 
In several cases, we can recover the peak density at high accuracy ($\Delta \log_{10} \rho_{\rm max} \ll 1$). 
However, we emphasize that even order-of-magnitude estimates ($\Delta \log_{10} \rho_{\rm max} \lesssim 3$) will be enough to place stringent constraints on the DM models. 
$\Delta \log_{10} \rho_{\rm max} \lesssim 3$ is satisfied in the range $M_{\rm BH}\in [10^3, 10^{4.5}]\rm M_\odot$ and $M_{\rm BH}\in [10^3, 10^6]\rm M_\odot$ for $\gamma=1$ and $\gamma=2$~(thick solid and dashed lines, respectively). Beyond that, we consider the event ``unmeasurable.'' 
We apply Gaussian filter for data visualization purposes. 
}
 \label{fig:densityrecovery}
\end{figure}

Figure~\ref{fig:densityrecovery} shows the expected $\rho_{\rm peak}$ recovery uncertainty for the above setup. The peak density can be recovered at a high accuracy ($\Delta \log_{10} \rho_{\rm peak} \ll 1$) for large ranges of $\rho_{\rm peak}$ and $M_{\rm BH}$. However, we emphasize (and show below) even order-of-magnitude estimates ($\Delta \log_{10} \rho_{\rm peak} \lesssim 3$) will be enough for us to place stringent bounds on the DM models. Specifically, we find that $\Delta \log_{10} \rho_{\rm peak} \lesssim 3$ at BH masses $M_{\rm BH}\in [10^3, 10^{4.5}]\rm M_\odot$ and $M_{\rm BH}\in [10^3, 10^6]$ for $\gamma = 1$ and $\gamma = 2$, respectively. We consider the cases where $\Delta \log_{10} \rho_{\rm peak} > 3$, as ``unmeasurable.'' At all the considered values, the spike leaves a noticeable orbital shift on the compact object's trajectory and therefore is measurable in gravitational wave emission. Below $10^{3}M_{\odot}$, the detector sensitivity deteriorate rapidly~\cite{gair2017prospects}.

We account for degeneracies between the 2.5 PN binary parameters and the DM spike parameters by the FIM approach. 
More accurate waveforms introduce higher order effects. 
In principle, these corrections could be degenerate with the gravitational drag induced by the spike. 
However, we expect the degeneracies to be small: 
the spike introduces a slow cumulative phase shift due to gravitational drag, which is quite distinct from the higher order binary effects such as spin precession. 

We note that our results do not strongly depend on the final spike index, but is most sensitive to $\rho_{\rm peak}$.
For simplicity, we fix the SNR and $q$. A more realistic, larger SNR would lower the measurable line~(in Fig.~\ref{fig:densityrecovery}), e.g., by a factor of a few for SNR = 100. The detectability also improves by a factor of a few if we choose $q = 10^{-3}$, and vice versa for $q = 10^{-5}$.  However, we note that our waveform approximation likely breaks down at smaller values of $q$ ($q\ll 10^{-4}$). We leave the more accurate and detailed exploration of the parameter space with for future work. 

Finally, we note that in our default DM spike model, the final density index is always $\alpha>2.25$.  If the astrophysical spike disruption effects are important, the slope could be flattened, perhaps ending with a shallow case $\alpha\simeq 1.5$ or an intermediate case $\alpha\simeq 1.8$~\cite{Fields:2014pia}.  Even in the shallow case, the spike can still be detectable at $\{M_{\rm BH} = 10^3\,M_{\odot},\gamma = 2\}$; while in the intermediate case, the spike is detectable at $\{M_{\rm BH} = 10^3\,M_{\odot},\gamma \simeq 1.6\}$ or $\{M_{\rm BH} \leq 10^4\,M_{\odot},\gamma = 2\}$. \emph{The EMRI GW probe is thus sensitive to a broad range of spike parameters. }

\noindent\textbf{\underline{EMRI detection rates}} 
Studies of the expected EMRI rate with self-consistent BH formation and evolution models suggest $\cal O$(100--1000) EMRI events throughout the lifetime of LISA~\cite{Babak:2017tow}. Importantly, these models suggest a mildly negative BH mass function, thus favors events with lighter BHs. DM spikes around lighter BHs are denser and thus easier to detect (Fig.~\ref{fig:densityrecovery}). Also, light BHs are less likely to suffer from major mergers and are thus more likely to retain the DM spikes~\cite{zhao2005dark, bertone2005new,eda2013new, eda2015gravitational}.

\section{Dark matter tests with EMRI}
\noindent
If DM consists of ultralight bosons, light fermions, self-annihilating particles, or primordial BHs~(PBHs), the DM spike density would be affected. Measuring the DM spike with EMRI could test these scenarios \emph{simultaneously}. 

\noindent\textbf{\underline{Ultralight bosonic DM:}} 
Heisenberg's uncertainty principle causes the light bosons (with mass $\mu_{s}$) to dilute the DM spikes and set a theoretical upper bound on their density. Such theoretical upper bound on density allows us to constrain the light boson mass by spike observations. 

Consider an initial scalar field $\psi_i$ surrounding the center of a galaxy without a black hole. Since the boson mass is light, it forms a BEC (i.e., it is in its ground state; $\psi_i = \psi_i^{\rm ground}$ for some initial Hamiltonian)~\cite{hu2000fuzzy,urena2002supermassive,lundgren2010lukewarm,cruz2011scalar,wavedarkmatter,zhang2018ultralight}. 

A black hole then grows adiabatically in the center, evolving the surrounding scalar field. 
The growth takes place on much larger timescale than the scalar field cycle\footnote{The scalar field oscillation time-scale is $\tau \sim \mu_s^{-1} \lesssim 100$ yr $\ll 10^6$ yrs when $\mu_s \gtrsim 10^{25}$ eV~\cite{detweiler1980klein,arvanitaki2010string,brito2015superradiance,brito2015black}}, thus adiabatic theorem~\cite{sakurai1967advanced, nenciu1980adiabatic,mostafazadeh1996adiabatic} applies: 
The final state of the scalar field $\psi_f$, after growth, will be in the ground state of the final Hamiltonian, which is approximately the black hole Hamiltonian near the center~\cite{urena2002supermassive}. 

Thus, the final state after growth is the ground state of a black hole/boson system, which has an exact solution in the very light boson mass limit ($M_{\rm BH} \mu_s \ll 1$)~\cite{detweiler1980klein}. 
The density of the ground state is~\cite{detweiler1980klein,baumann2019probing}:
\begin{equation}\label{eq:scalarfield}
 \rho_s(\mu_{s}, M_{\rm BH},r, \theta, M_s) \simeq  \frac{M_s}{64 \pi \mu_s} M_{\rm BH}^5 \mu_s^{11} r^2 e^{-M_{\rm BH} \mu_s^2 r} \sin^2\theta,
\end{equation}
where $r$ is the radius, $M_s=\int \rho r^2 dr d\Omega$ is the cloud mass, and the density has been expanded in leading order of $r^{-1}$. 
Note that we assume complex scalar fields, but real fields share similar predictions~\cite{Dolan:2007mj,arvanitaki2010string,arvanitaki2015discovering,brito2015superradiance}.

\begin{figure*}
 \includegraphics[width=\textwidth]{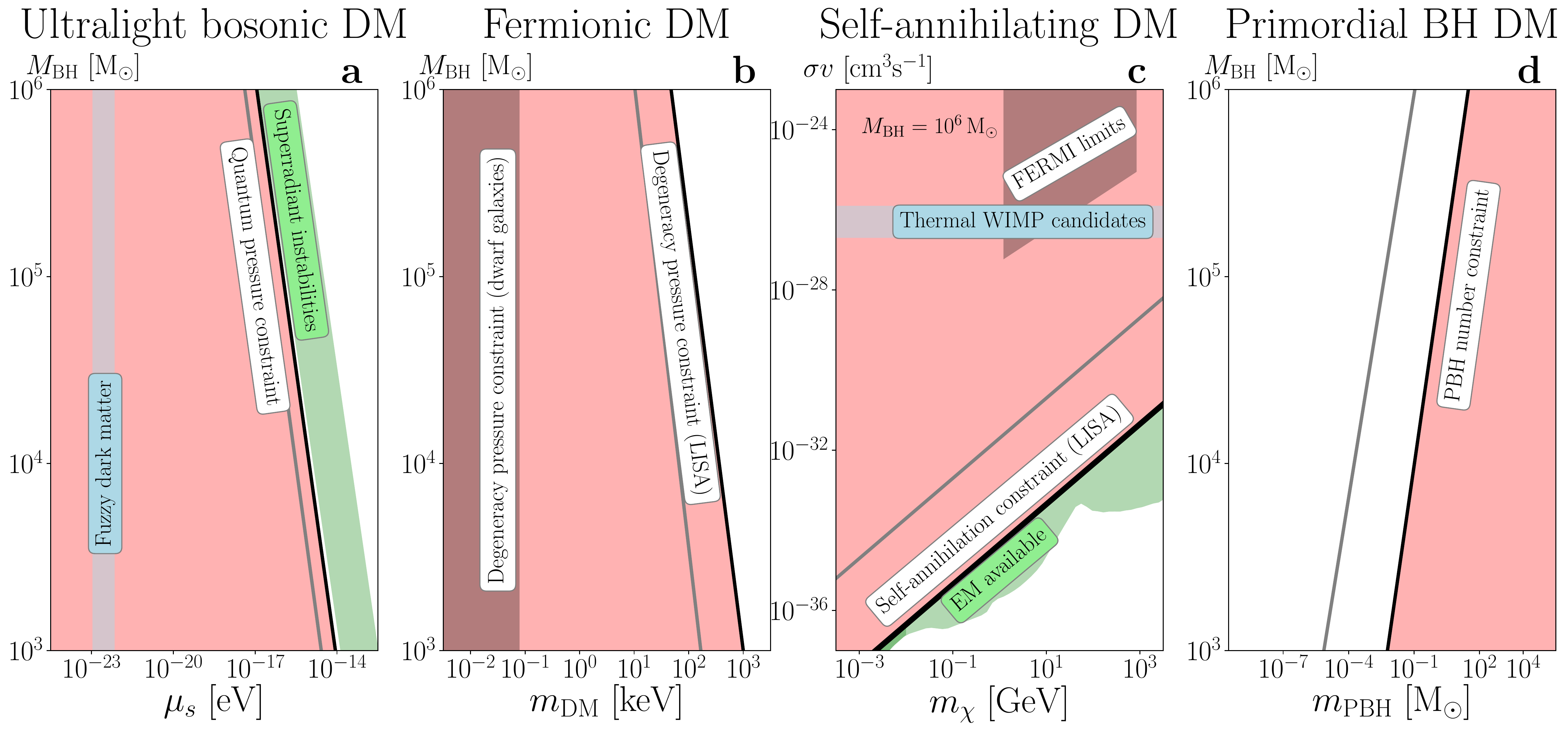}
 \caption{New constraints~(red shaded regions) on DM models if a DM spike is detected with an EMRI. For ultralight bosons~(panel \textbf{a}), fermionic DM~(panel \textbf{b}), and PBH DM~(panel \textbf{d}), we exclude a region of the DM particle/PBH mass. The constraints depend on the mass of the detected central black hole, $M_{\rm BH}$ (the DM spike profile is uniquely predicted for given $M_{\rm BH}$ using the $M$--$\sigma$ relation).  For self-annihilating DM~(panel \textbf{c}), the constraint is on the cross section-DM mass plane, assuming $M_{\rm BH}=10^6\, \text{M}_\odot$. 
 If ultralight bosons exist in the $m_{\rm DM} \in [10^{-17},10^{-14}]$ eV range, they could be identified through superradiant-induced clouds (see Ref.~\cite{hannuksela2019probing}; panel \textbf{a}, green region). 
 If the EMRI event is sufficiently nearby~($\simeq90\, \text{Mpc}$), as expected of the closest EMRIs~\cite{gair2009probing}, electromagnetic counterparts from DM annihilation may be possible in some optimistic cases~(panel \textbf{c}, green region).  Previous lower limits~(gray regions) on fermionic DM and upper limits on DM annihilation cross section are from Refs.~\cite{DiPaolo:2017geq, leane2018gev}.  For all panels, the thick solid lines and thin gray lines correspond to $\gamma = 2$ and $\gamma = 1$ initial DM halo slopes, respectively. See text and Appendix for details.}
\label{fig:constraintscollab}
\end{figure*}

The cloud mass must be smaller than the mass inside the influence radius ($M_s \lesssim M_s^{\rm max} \sim 2M_{\rm BH}$~\cite{Merritt:2003qc}).
This translates to constraints on the maximal observed density $\rho_{\rm obs}\lesssim \rho_{\rm max}= \rho_s(\mu, M_{\rm BH}, 2/(M_{\rm BH} \mu_s^2), \pi/2, M_s^{\rm max})$, which yields 
\begin{equation}\label{eq:bosonconstraint}
\begin{split}
 \mu_s \gtrsim & 10^{-17} {\, \rm eV} \left( \frac{\rho_{\rm obs}}{10^{20} \, \rm GeV/cm^3} \frac{2M_{\rm BH}}{M_s^{\rm max}}\right)^{1/6} \left( \frac{10^6 \, \rm M_\odot}{M_{\rm BH}} \right)^{2/3}.  
\end{split}
\end{equation}
Consequently, we could disfavour, for example, the fuzzy/wave DM candidate in the $\mu_s \in [10^{-23}, 10^{-21}]\, \rm eV$ range~\cite{wavedarkmatter,zhang2018ultralight}, proposed as a solution to many cosmological problems (Fig.~\ref{fig:constraintscollab}, panel \textbf{a}).

We note that if stars or compact objects of large mass are present near the BH and in sufficient abundance, it is possible that the boson cloud ground state (Eq.~\ref{eq:scalarfield}) could mix with higher order states or collapse back to the BH~\cite{baumann2019probing,berti2019ultralight}. 
In this case, the boson cloud will no longer reside in the ground state, but either does not exist or resides a higher state. 
However, this is not an issue: Higher modes are spread across an even \emph{larger} volume and therefore predict smaller densities~\cite{detweiler1980klein}, making our estimate conservative.

Note also that when the boson mass is larger\footnote{In the LISA BH range $M_{\rm BH}\in[10^3, 10^6] \, \rm M_\odot$ would have such "matching" bosons in the $\mu_s \in [10^{-17}, 10^{-14}] \, \rm eV$ range.}, superradiance can create bosonic clouds~\cite{brito2015black,brito2015superradiance}. 
These clouds may allow one to verify the existence of the bosons by this same EMRI measurement as shown in Ref.~\cite{hannuksela2019probing} (we show the target region in Figure~\ref{fig:constraintscollab}, panel \textbf{a}, green area, for completeness).

\noindent\textbf{\underline{Fermionic DM:}} If DM is made of fermions, such as sterile neutrinos, they occupy a finite phase space volume due to the exclusion principle. Thus, a measurement of the mass density can be used to derive robust limits on fermionic DM mass.

Consider a system of a degenerate fermionic DM, the Fermi velocity is
\begin{equation}
 v_{F} = \left( \frac{6\pi^{2}\hbar^{3} \rho}{m_{\rm DM}^{4} g} \right)^{1/3} \, .
\end{equation}
For the density spike to be stable, the Fermi velocity must be less than the escape velocity of the BH plus DM spike system
\begin{equation}
 v_{F} \leq v_{esc} \equiv \sqrt{ \frac{2 G \left( M_{\rm BH} + M_{\chi} \right)}{R} } \simeq \sqrt{ \frac{2 G M_{\rm BH}}{R} }  \, .
\end{equation}
This translates to a lower bound on the fermionic DM mass, given an observation of density $\rho_{\rm obs}$,
\begin{equation} \label{eq:fermionconstraint}
 m_{\rm DM} \gtrsim 30\,{\rm keV} \left( \frac{\rho_{\rm obs}}{10^{20}\,{\rm GeV/cm^{3}}} \frac{2}{g} \right)^{1/4} \left(\frac{R}{20 M_{\rm BH}} \right)^{3/8}.
\end{equation}
Thus, a detection of the DM spike will significantly improve existing fermioninc DM constraint~\cite{DiPaolo:2017geq} by more than 2 orders of magnitude.
While our result only depends on the measured density, we express the constraint in terms of $M_{\rm BH}$ using our reference DM spike model for consistency, as shown in Fig.~\ref{fig:constraintscollab} panel \textbf{a}.
This result is robust, and does not depend on the initial phase-space density distribution~\cite{Tremaine:1979we, Horiuchi:2013noa}. It will also close the $\nu$MSM sterile neutrino DM window~\cite{Canetti:2012vf, Canetti:2012kh} without relying on X-ray searches~\cite{Bulbul:2014sua, Boyarsky:2014jta, Ng:2015gfa, Speckhard:2015eva, Powell:2016zbo, Perez:2016tcq, Ng:2019gch}.

\noindent\textbf{\underline{Self-annihilating DM:}} 
Thermally produced weakly interacting massive particles is a popular generic DM candidate, and predicts a definite annihilation cross section~($\sim10^{-26} \, \rm cm^3 s^{-1}$ for s-wave, $2\rightarrow 2$) that is required to produced the observed DM abundance~\cite{Steigman:2012nb}. The self-annihilation would also rapidly smooth out the DM spike into  ``annihilation plateaus''~\cite{sadeghian_dark_2013}.
Thus, observation of the DM spike places a limit on the annihilation cross section $\sigma v$.

We approximate the annihilation plateau as a flat core, $\rho_{\rm core} = m_\chi/(\sigma v t_{\rm BH})$~\cite{gondolo_dark_1999}~(see also~\cite{vasiliev2007dark, vasiliev2008dark,shapiro_weak_2016}). Taking conservatively the age of the BH to be $t_{\text{BH}}\gtrsim 10^6$ years~(much less than the mean age of galaxies or stars, $\sim 10^{10} \, \rm yr$), an EMRI DM spike measurement sets a upper limit on the \emph{total} annihilation cross section,
\begin{equation} \label{eq:crosssectionconstraint}
 \begin{split}
 \sigma v &\lesssim 3.17 \times 10^{-32} \, \text{cm}^3 \text{s}^{-1}  \left( \frac{m_{\chi}}{100 \, \text{GeV}} \right) \\ 
          & \times \left( \frac{10^{20} \, \text{GeV}/\text{cm}^{3}}{\rho_{\rm obs}} \right) \left( \frac{10^6 \, \text{yr}}{t_{\text{BH}}} \right). 
 \end{split}
\end{equation}
Thus, any EMRI DM spike measurement will be in strong tension with the simplest thermal relic DM hypothesis~(Fig.~\ref{fig:constraintscollab}, panel \textbf{c}), currently an open window between $20\,{\rm GeV}<m_{\chi}<100\,{\rm TeV}$~\cite{leane2018gev}.
We emphasize this is the total cross section, thus includes the difficult-to-probe neutrino channels.

For other cases~(p-wave annihilation, non-thermal models, and others), the cross-section could be significantly lower~\cite{Shelton:2015aqa}.  In such scenarios, the EMRI event could have a persistent electromagnetic counterpart due to DM annihilation.
We find that in the optimistic scenario, where $\gamma=2$, the BH is heavy $M_{\rm BH}\sim 10^6 \, \rm M_\odot$ and nearby $D\sim 90 \, \rm Mpc$, the electromagnetic counterpart is detectable with e-ASTROGAM/Fermi/CTA~(see the Appendix for details), as shown in Fig.~\ref{fig:constraintscollab}. However, within such a small volume, the expected number of EMRI events is only of order one~\cite{gair2009probing}.  Thus, to see the electromagnetic counterpart, the fraction of halos hosting spikes must be high, the DM spike must be relatively young, and the event must be nearby.

\noindent\textbf{\underline{Primordial black hole DM}} We consider the case that PBHs dominates cosmic DM density~(as long as the non-PBH component of DM cannot mimic a spike as described above.) 
If a DM spike is measured with EMRI, there must be at least one PBH~($N\geq1$) in the probed volume~($8M_{\rm BH}<r<300M_{\rm BH}$). The mass of the of the PBH,$m_{\rm PBH}$, must then satisfy 
\begin{equation}\label{eq:PBHConstraint}
 N m_{\rm PBH} \leq \int_{8M_{\rm BH}}^{300M_{\rm BH}} \rho_{\rm obs}(r) d^{3}r \, . 
\end{equation}
Consequently, the PBH mass range could be constrained to $m_{\rm PBH} \lesssim 10^{-7}\, \rm M_\odot$ ($\gamma =1$, Fig.~\ref{fig:constraintscollab}, panel \textbf{d}). 
This simple PBH number argument offers an independent constraint on the PBHs, complementary to various existing considerations~(e.g., Refs.~\cite{capela2013constraints,Blum:2016cjs, Boudaud:2018hqb, DeRocco:2019fjq, Laha:2019ssq, Ballesteros:2019exr}). 

We note that the above $N=1$ constraint is exceptionally conservative. In the case of a spike detection, for the PBH DM to mimic the dynamical friction effect, the spike must have a large amount of PBHs $N \gg 1$, thus leading to more stringent constraints. We leave the exploration of this effect for the future. 

In principle, one could also combine these spike measurements with ground-based detectors that observe PBH mergers within the spikes. 
Unfortunately, the fraction of mergers within a single halo is $N_{\rm sp} \lesssim 10^{-2} \, \text{yr}^{-1}$ (Ref.~\cite{nishikawa2017primordial}]), and thus constraining PBHs by aid of ground-based detectors will be difficult. 
We note that PBHs themselves also act as EMRIs, which could offer another channel for their detections~\cite{kuhnel2018waves}.

\section{Discussion and Conclusion}
\noindent

We have shown that, in the near future, EMRI GW measurement with space interferometry experiments could place strong constraints on DM models across the particle landscape. The EMRI GW emission will provide model-independent (purely gravitational) tests for DM spikes, which is a prediction of cold collisionless DM models. If such spikes are detected, they will be the most compact DM structures ever known. 

These compact structures are incompatible with several popular DM models, such as ultra-light bosons, keV fermions, self-annihilating DM, and PBHs. 
Note that in the DM parameter space we consider here, the DM spike will always be flattened due to the intrinsic DM particle properties. On the other hand, as discussed earlier, astrophysical effects may also partially flatten the spike. A large number of expected EMRIs ($\sim100-1000$) is thus extremely advantageous; it allows EMRIs to probe BH systems under various astrophysical conditions and with variable merger histories, reducing the chance of non-detection of the spike due to astrophysical effects. 

If no spikes are detected, then, unfortunately, no conclusions can be drawn immediately.
One possibility is that astrophysical processes~(stellar heating, mergers, etc~\cite{ullio_dark-matter_2001,merritt_dark_2002,bertone_time-dependent_2005,bertone2005dark}) had destroyed the spikes. The processes must then be common, robust, and applicable to different BHs masses and galaxy properties. Follow-up and detailed astronomical observations of the EMRI events~(e.g., ~\cite{Akiyama:2019cqa}) are then necessary to identify the astrophysical mechanism. Indeed, galaxies with large EMRI rates may tend to have stellar cusps, which could potentially negatively bias their likelihood of hosting DM spikes~\cite{Babak:2017tow}. The properties of DM might also cause the smoothing of the spike.  In these cases, the GW observation itself may already contain smoking-gun signatures of the particle DM~(e.g.,\cite{hannuksela2019probing}). Otherwise, independent probes are needed to pinpoint the particle effect~(e.g., gamma-ray searches or others~\cite{Davoudiasl:2019nlo, Bar:2019pnz}).  

Here we provide the principal methodologies for DM searches by EMRIs. To fully realize the search potential, realistic waveforms~\cite{Babak:2006uv,Chua:2017ujo,Babak:2017tow} that can capture the effect of DM are required. These waveforms would need to be suitable for the extreme gravity and mass ratio that we consider here, and should also be generic in their ability to distinguish DM distributions. The form of the spike distribution can in principle be quite generic, and therefore more flexible waveforms, specialized towards EMRIs, should be developed before they can be applied to realistic LISA data analysis. 

Despite decades of searches, the identity of DM is still elusive. 
If DM spikes are detected, they could provide much-needed sensitivity to revolutionize the landscape of DM searches.

\section*{\label{sec:acknowledgements} Acknowledgments}

We thank Emanuele Berti, Richard Brito, Kaze W. K. Wong for their work on a very closely related project, Yifan Wang for discussion on primordial black holes and Ming-Chung Chu and Po Kin Leung for their supervision on a project on dark matter spike modeling. 
We also thank John F. Beacom for his helpful comments, Richard Brito for his advice on ultralight bosons, and Kaze W. K. Wong for his comments on data-analysis. 
O.A.H. is supported by the Hong Kong Ph.D. Fellowship Scheme (HKPFS) issued by the Research Grants Council (RGC) of Hong Kong.
K.C.Y.N. is supported by a Croucher Fellowship and a Benoziyo Fellowship.
T.G.F.L. was partially supported by grants from the Research Grants Council of the Hong Kong (Project No. CUHK14310816, CUHK24304317 and CUHK14306218), The Croucher Foundation in Hong Kong, and the Research Committee of the Chinese University of Hong Kong.

\appendix
\section{Appendix}
\subsection{Relating the dark matter spike profile to the host black hole} \label{app:densityspikeconstruction}

When a BH surrounded by $\rho_i(r) \simeq \rho_0(r_0/r)^\gamma$ DM profile grows adiabatically, it gravitationally concentrates the surrounding matter around it to a cuspy profile usually referred to as a ``DM spike''~\cite{gondolo_dark_1999}.
The DM density spike can be modelled generally as~\cite{nishikawa2017primordial}
\begin{equation}\label{app:eq:spike}
 \rho_{\rm sp}(r) = \rho_R \left( 1-\frac{8M_{\rm BH}}{r} \right)^3 \left(\frac{r}{R_{\rm sp}}\right)^{-\alpha},
\end{equation}
where $\rho_R = \rho_0 (R_{\rm sp}/r_0)^{-\gamma}$, $R_{\rm sp} = a_\gamma r_0 (M_{\rm BH}/(\rho_0 r_0^3))^{1/(3-\gamma)}$, $a_\gamma$ is a numerical fit with $a_1\simeq 0.1$ and $a_2\simeq0.02$~\cite{gondolo_dark_1999} and $\gamma_{\rm sp}=(9-2\gamma)/(4-\gamma)$.
The parameters $\rho_0$ and $r_0$ can be related to the initial density profile outside the spike, taken to be the NFW profile, via the BH mass through the $M-\sigma$ relationship, the one-dimensional halo velocity dispersion and the virial mass relations ~\cite{nishikawa2017primordial}
\begin{align}\label{eq:constraintsapp1}
 \log_{10}(M_{\rm BH}/{\rm M_\odot}) &= a + b \log_{10}(\sigma / {\rm 200 km s^{-1}}),\\
 \sigma^2                   &= \frac{4 \pi G \rho_0 r_0^2 g(c_m)}{c_m},\\
 M_{\rm vir} &= 4 \pi \rho_0 r_0^3 g(c(M_{\rm vir})), \, \label{eq:massintegral} \\
 M_{\rm vir} &= \frac{4}{3}\pi R_{\rm vir}^{3}\Delta \rho_{c} \, ,
\end{align}
where $a=8.12$ and $b=4.24$ are empirically determined parameters, $\sigma$ is the one-dimensional halo velocity dispersion of the galaxy, $g(x)=\log(1-x)-x/(1+x)$, which is obtain by the mass the mass integral of the NFW profile~(Eq.~\ref{eq:massintegral}), $c_m=2.16$,
$\Delta\simeq 200$, and $\rho_{c}$ is the critical density. The concentration~($c = R_{vir}/r_{0}$) relation is~\cite{sanchez2014flattening}
\begin{equation}~\label{eq:constraintsapp2}
 \begin{split}
 c(M_{\rm vir})&= c_5 \log ^5\left(\frac{h_0 M_{\rm vir}}{{\rm M_\odot}}\right)+c_4 \log ^4\left(\frac{h_0 M_{{\rm vir}}}{{\rm M_\odot}}\right)\\
               &+c_3 \log ^3\left(\frac{h_0 M_{{\rm vir}}}{{\rm M_\odot}}\right)+c_2 \log ^2\left(\frac{h_0 M_{{\rm vir}}}{{\rm M_\odot}}\right)\\
               &+c_1 \log \left(\frac{h_0 M_{{\rm vir}}}{{\rm M_\odot}}\right)+c_0,
 \end{split}
\end{equation}
where $h_0=0.67$, $c_0=37.5153$, $c_1=-1.5093$, $c_2=1.636 \times 10^{-2}$, $c_3=3.66 \times 10^{-4}$, $c_4=-2.89237 \times 10^{-5}$ and $c_5=5.32 \times 10^{-7}$.
Using Eqs.~\ref{eq:constraintsapp1}-\ref{eq:constraintsapp2}, we use the $M-\sigma$ relation to connect $M_{\rm BH}$ and $M_{\rm vir}$
\begin{equation}
    \sigma_{v}^{2}(M_{\rm BH}) = \frac{4 \pi}{3}G \Delta \rho_{c} \frac{g(c_{\rm m})c }{g(c)c_{\rm m}} \left( \frac{3 M_{\rm vir}}{4 \pi \Delta \rho_{c}} \right)^{2/3} \, .
\end{equation}
We can then solve for $M_{\rm vir}$ numerically for each $M_{\rm BH}$, and subsequently obtain $\rho_R$ and $R_{\rm sp}$. 

We note that the inner part of the density profile~(roughly outside the spike) is not well constrained~\cite{pato2015dynamical}.
We, therefore, choose two reference values for our profile; $\gamma=1$, which corresponds to the NFW profile. In some cases~\cite{gnedin2004response}, the density slope could be steeper.  We thus consider $\gamma=2$ to be the most optimistic scenario. 

\subsection{Electromagnetic Counterparts from DM annihilations} \label{app:emcounterpart}

For DM annihilation in the DM spike, the differential gamma-ray flux is 
\begin{equation}
    \frac{dF}{dE} = \frac{\sigma v}{2 m_{\chi}^{2}} \frac{\int \rho_{\rm sp}^{2}dV}{4 \pi D^{2}} \frac{dN}{dE} \, ,
\end{equation}
where $\sigma v$ is the annihilation cross section, $m_{\chi}$ is the DM mass, $D$ is the distance to the source, and $dN/dE$ is the gamma-ray spectrum per annihilation.  For simplicity, we assume 100\% $\chi\chi\rightarrow \gamma\gamma$ below 10\,GeV and $\chi\chi\rightarrow \tau\tau$ above.  For the $\gamma\gamma$ channel, we assume delta function spectrum convolved with 10\% energy resolution, a typical detector resolution for gamma-ray detectors.  For the $\tau\tau$ channel, we obtain the spectrum from PPPC4DMID~\cite{Cirelli:2010xx}. 

For detecting the gamma-ray flux, we consider the joint differential sensitivity in 1\,MeV--10\,TeV from 1 year of e-ASTROGAM~\cite{DeAngelis:2017gra}, 10 years of Fermi~\cite{fermi_sensitivity}, and 50 hours of CTA~\cite{cta_sensitivity}. 

We obtain the sensitivity~(green band, Fig.~\ref{fig:constraintscollab}, panel \textbf{c}) requiring the DM flux to touch the detector sensitivity described above.  This is conservative, as in practice sharp spectra from DM annihilation typically can improve the sensitivity by a factor of a few, after taking into account the smooth astrophysical and detector backgrounds.

\subsection{Parameter inference of the dark matter distribution} \label{app:densityspikeinference}

Our GW waveform strain is of the form
\begin{equation}
 h(f) = \frac{\sqrt{3}}{2}  A f^{-7/6} e^{i(\psi(f) + \psi_{\rm DM}(f))},
\end{equation}
where $f$ is the GW frequency, $A=\mathcal{M}_c^{5/6}/[(30^{1/2} \pi ^{2/3}) d_L]$ is the amplitude with $\mathcal{M}_c$ being the chirp mass. 
The phase of the GW $\psi(f)$ is approximated up to 2.5 post-Newtonian order as in~\cite{Berti:2004bd}.
\begin{equation}
\begin{split}
 \psi(f) = & 2 \pi  f t_c-\phi_c  \\ 
 &+ \frac{3}{128} X^{-5/3} \left(1 + \left(\frac{55 \eta }{9}+\frac{3715}{756}\right) \eta ^{-2/5} X^{2/3}  \right. \\
 &\left. +\left(\frac{3085 \eta ^2}{72}+\frac{27145 \eta }{504}+ \frac{15293365}{508032}\right) \eta ^{-4/5} X^{4/5} \right. \\
 &\left. +4 \beta  \eta ^{-3/5} X-16 \pi  \eta ^{-3/5} X  -10 \eta ^{-4/5} \sigma  X^{4/3}\right),
\end{split}
\end{equation}
where $t_c$ and $\phi_c$ are the time and phase of coalescence, $X=\pi \mathcal{M}_c f$, $\eta$ is the symmetric mass ratio and ($\beta$, $\sigma$) are the binary spin parameters. 
In all our calculations, we assume angle-averaged LISA antenna pattern functions.

We construct the phase shift introduced by a DM spike $\psi_{\rm DM}(f)$ similarly to Ref.~\cite{eda2015gravitational}. 
We choose the DM spike of the form of
\begin{equation}
 \rho_{\rm dm}(r) = \rho_{\rm peak} \left( \frac{r}{20 M_{\rm BH}}\right)^{-\alpha},
\end{equation}
where $\rho_{\rm peak}$, $\alpha$ are the spike parameters. 
The spike introduces a gravitational pull and a dynamical friction onto the binary orbit~\cite{macedo2013into,eda2015gravitational} 
\begin{equation} \label{eq:energybalance}
 \frac{dE_{\rm orbit}}{dt} + \frac{dE_{\rm gw}}{dt} + \frac{dE_{\rm DF}}{dt} = 0,
\end{equation}
where
\begin{equation}
\begin{split}
 \frac{dE_{\rm orbit}}{dt} &= \frac{d}{dt} \left( \frac{1}{2} \mu  v^2+\mu\Phi (r) \right),\\
 \frac{dE_{\rm gw}}{dt}    &= \frac{32 \mu^2 r^4 \omega^6}{5},\\
 \frac{dE_{\rm DF}}{dt}    &= \frac{4 \pi  \mu ^2 \log \Lambda \rho_{\rm dm}(r)}{v(r)},
\end{split}
\end{equation}
with $v$ and $\omega$ being the velocity and orbital frequency of the smaller compact object, $\log \Lambda \sim 3 + \log [(10^{-4}/q) (20M_{\rm BH}/r)]$ is the Coulomb logarithm~\cite{binneytremaine} which depends weakly on the mass ratio $q$ and orbital radius $r$~\cite{binneytremaine}, $\mu$ the reduced mass of the system (which we take to be approximately the component mass),  and $\Phi(r)$ is the Newtonian potential of the spike and the black hole.

The most dominant effect that the collisionless dark matter introduces is \emph{dynamical friction}; the accretion onto the black hole~\cite{yue2018gravitational} (which we ignore) and gravitational pull are both sub-dominant~\cite{barausse2015environmental,yue2018gravitational}.  
We assume a constant value of the Coulomb logarithm throughout the inspiral, as in~\cite{eda2015gravitational}, computed at $r=20M_{\rm BH}$ and given mass ratio $q$.

Dynamical friction stems from the gravitational pull by high-density "tails" that form behind the compact object in its wake.
We note that ultralight bosons could, in principle, inhibit the formation of these tails due to quantum effects, similarly as they inhibit the formation of high-density peaks within galaxies~\cite{wavedarkmatter}. 
However, the precise framework to model the gravitational effects by light bosons (or more generally scalar fields) is still being worked out~\cite{kesden2005gravitational,becerril2006geodesics,diemer2013geodesic,macedo2013astrophysical,macedo2013into,brihaye2014charged,grandclement2014models,bovskovic2018motion,ferreira2019scalar}. 
If the tails and thus dynamical friction are indeed inhibited, then a measurement of the spike would imply even \emph{stronger} constraints on the ultralight bosons. 

We obtain the rate of change of orbital radius $r'(t)$ from the energy evolution equation. 
Inverting the orbital frequency yields $r(f)$~\cite{eda2015gravitational}, which can be translated to the total GW phase shift using the stationary phase approximation~\cite{Maggiore:1900zz}.

The mapping between the orbital radius and the GW frequency may be solved by inverting the following relation for $r$
\begin{equation}
 \omega(f) = \pi f = \sqrt{\frac{\Phi '(r)}{\mu r}},
\end{equation}
and expanding to first order in matter effects (Ref.~\cite{eda2015gravitational}). 

We then estimate the parameter recovery using standard Fisher Information Matrix (FIM) approach, which applies in the high signal-to-noise ratio limit (as implemented in~\cite{maggiore,Berti:2004bd}), and assume LISA power spectral density~\cite{Cornish:2018dyw}.  
The recovered parameters are $\vec{\theta} \in \{ A, \phi_c, t_c, \log \mathcal{M}_c, \log \eta, \beta, \sigma, \rho_0, \alpha \}$.

Let us note a common issue with the FIM approach: If the FIM is high-dimensional, the FIM inversion can become unstable for specific sets of parameters~\cite{Berti:2004bd,vallisneri2008use}. 
In Fig.~\ref{fig:densityrecovery}, we exclude these unstable sets of parameters in our analysis and replace them by values interpolated from the nearest stable points.

\section{Adiabatic theorem}

Consider wave function of two non-degenerate states in the non-relativistic limit, such that we can separate the wave function of the bosonic cloud in two components (similarly to ~\cite{baumann2019probing,berti2019ultralight})
\begin{equation}
\ket{\psi(t)} = c_n(t) e^{i \theta_n(t)}\ket{\psi_n(t)} + c_m(t) e^{i \theta_m(t)} \ket{\psi_m(t)}\,,
\end{equation}
such that $|c_n(t)|^2 + |c_m(t)|^2=1$ and $c_n(0) = 1$ and $c_m(0)=0$. 
The exponential phase is introduced for convenience and satisfies $\theta_n(t) = - \int_0^t \omega_n(t^\prime) dt^\prime$. 

Inserting this into the Schr\"odinger equation, we find
\begin{equation}
 \sum_{j=n,m} e^{i \theta_j(t)}\left[ \dot{c}_j(t) \ket{\psi_j(t)} + c_j(t) \partial_t \ket{\psi_j(t)} \right] = 0.
\end{equation}
Taking inner product with respect to $\bra{m}$, we get
\begin{equation}
 \dot{c}_m(t) = - \sum_{j=n,m} c_j(t) e^{i[\theta_j(t)-\theta_m(t)]} \bra{\psi_m(t)}\left[ \partial_t \ket{\psi_j(t)} \right].
\end{equation}
Inserting the following generic result for Schr\"odinger equation
\begin{equation}
 \bra{\psi_m(t)} \dot{H} \ket{\psi_n(t)} = [ \omega_n(t) - \omega_m(t) ] \bra{\psi_m(t)} \partial_t \ket{\psi_j(t)},
\end{equation}
we get
\begin{equation}
\begin{split}
 \dot{c}_m(t) = &- c_m(t) \bra{\psi_m(t)} \partial_t \ket{\psi_m(t)}\,\\ 
 &- c_n(t) e^{i ( \theta_n(t) - \theta_m(t))} \frac{\bra{\psi_m(t)} \dot{H} \ket{\psi_n(t)}}{\omega_n-\omega_m}\,,
\end{split}
\end{equation}
where the second term can be neglected whenever~\cite[e.g.][]{sakurai1967advanced}
\begin{equation}
    \frac{\bra{\psi_m(t)} \dot{H} \ket{\psi_n(t)}}{\omega_n-\omega_m} \equiv \frac{1}{\tau} \ll \bra{\psi_m(t)} \left[ \partial \ket{\psi_m(t)} \right] \sim \omega_m.
\end{equation}
I.e., the Hamiltonian changes over a much longer period of time than the oscillation frequency of the system, which for given ($nlm$) state is~\cite{baumann2019probing, berti2019ultralight}
\begin{equation}\label{eigenfreqs}
\omega_{n\ell m}\simeq \mu\left(1-\frac{\alpha^2}{2n^2}+\delta\omega_{n\ell m}\right)\,,
\end{equation}
where $\delta \omega_{n\ell m}$ denote higher-order corrections, that (up to
fifth-order in $\alpha$) are given by~\cite{baumann2019probing,berti2019ultralight}
\begin{equation}\label{eigenfreqs_2}
\begin{split}
\delta\omega_{n\ell m} \simeq &\left(-\frac{\alpha^4}{8n^4}+\frac{(2\ell-3n+1)\alpha^4}{n^4(\ell+1/2)}\, \right.\\
                              &\left.+\frac{2\tilde{a}m\alpha^5}{n^3\ell(\ell+1/2)(\ell+1)}\right),
\end{split}
\end{equation}
where $\tilde{a}=a/M_{\rm BH}$ is the dimensionless spin. 
The ground state in this notation is $n=1$, $l=0$, $m=0$ state. 
To first order, the oscillation time-scale is
\begin{equation}
    \omega_{n\ell m}^{-1}\sim 2 {\, \rm yr} \left( \frac{\mu}{10^{-23} \, \rm eV} \right),
\end{equation}
which is much smaller than the black hole growth time-scale ($\sim 10^{6} \, \rm yr$). 
The adiabatic growth assumption is only violated at $\mu \ll 10^{-25} \, \rm eV$ mass range. 

As a cautionary note, the ground state is to a large degree degenerate with different $m$ modes: $m=-1,0,1$. 
This could potentially cause mode mixing of the stable ground mode with different (unstable) $m$ modes. 
However, we can safely neglect the mixing, as these states are both similarly diluted as the ground state. 
Hence, our estimate (Eq.~\ref{eq:scalarfield}) is conservative. 

\bibliographystyle{unsrt}

\end{document}